\documentclass[12pt,a4paper]{article}
\usepackage{amsmath}
\usepackage[dvips]{graphicx}
\newcommand{\bc}{\begin{center}}
\newcommand{\ec}{\end{center}}
\newcommand{\be}{\begin{equation}}
\newcommand{\ee}{\end{equation}} 
\newcommand{\ba}{\begin{eqnarray}}
\newcommand{\ea}{\end{eqnarray}}
 
\begin{document}
 
 \baselineskip 24pt
 \bc {\Large \bf Gravitational Self-Energy and Black Holes in Newtonian Physics}

G. Dillon\footnote{e-mail: dillon@ge.infn.it} \\ 
\baselineskip 16pt
{\it Dipartimento di Fisica, Universit\`a di Genova\\
INFN, Sezione di Genova} \ec

 \baselineskip 14pt
 \noindent{\large\bf Abstract:} A definition of a Newtonian black hole is possible which incorporates 
the mass-energy equivalence from special relativity. However, exploiting a spherical double shell model, 
it will be shown that the ensuing gravitational self-energy and mass renormalization prevent the 
formation of such an object.
 \vskip10pt
 \noindent PACS numbers: 04.40.-b, 04,50.kd 

\baselineskip 15pt 
\section{Introduction}
 The possible existence of a celestial object so massive to hold back even light with its gravity goes 
back to the end of '700 \cite{Mi,La}. According to Newtonian physics a spherically symmetric 
distribution of a mass $M$  inside a region of radius $R$  centered at the origin, yields the 
gravitational  potential (for $r\ge R$)
\be \Phi(r)=-G\frac{M}{r}
\label{1}\ee
Hence the energy of a test-particle $\delta m_0$ settled on its surface is
\be \delta U(R)=-G\frac{M\delta m_0}{R}
\label{2}\ee
If rays of light were constituted by a flux of tiny particles with a given kinetic energy (as was 
believed at the time) one would immediately get, from the conservation of the mechanical energy 
(kinetic+potential), the condition for the mass $M$ to be heavy enough to prevent light to escape from 
its surface. This condition defines a ``Newtonian black hole" (NBH).

It is a widespread opinion that an ``up-to-date" definition of a NBH is possible if one plugs Einstein's 
special relativity into Newtonian gravitation. Indeed, taking into account the mass-energy equivalence 
together with the inertial-gravitational mass equality, one may write for the total mass $M_t(R)$ of the 
system (heavy mass $M$ + test-particle $\delta m_0$ on its surface)
\be M_t(R)=M+\delta m_0+\delta U(R)/c^2=M+\delta m_0(1+ \Phi(R)/c^2)
\label{3}\ee
For a relativistic particle (\ref{3}) is supposed to hold as well, provided $\delta m_0c^2$ represents 
the full relativistic energy of the particle. For a photon it is: $\delta m_0=\hbar\omega/c^2$.  

Now if 
\be \delta U(R)=-\delta m_0 c^2
\label{4}\ee
one has from (\ref{3}) $M_t(R)=M$, i.e. the total energy of the system  with or without $\delta m_0$ is 
the same. This means, for example, that a photon, leaving the surface of that sphere, must spend its 
whole energy $\hbar \omega$ to get out from the gravitational field and will end its journey with a 
vanishingly small  frequency irrespective of the initial one. Therefore (\ref{4}) is the up-to-date 
condition for the existence of a NBH. A given mass $M_0$ confined  in a sphere of sufficiently small 
radius $R_0$:
\be R_0=GM_0/c^2
\label{5}\ee
leads to (\ref{4}).
If $R<R_0$ the meaning of $R_0$ is the maximum radial distance from where light cannot escape and 
corresponds to the so called ``event horizon" in the theory of the black holes in General Relativity 
(GR) \cite{t'}. Note that $R_0$ happens to be one-half of the Schwartzschild radius $R_S$. Anyhow the 
conceptual difference with the event horizon should be kept in mind because in GR the very structure of  
the space-time is drastically changed beyond $R_S$ and even the ``one-way passage"  (i.e. the fact that 
things are free to go inside $R_S$ but never to go outside) is unobtainable in Newtonian physics 
\cite{Ray}. 

However, if we take into account the mass-energy equivalence, we should also take into account the 
self-energy of the sphere. For example, in classical Newtonian physics, the gravitational energy of a 
simple  spherically symmetric shell of radius $R$ and mass $M_0$ turns out to be
\be U(R)=-G\frac{M_0^2}{2R}
\label{6}\ee
Such a binding energy (negative) is equivalent to a mass defect. Hence the mass of the shell will be 
different from $M_0$. In the following we shall refer to $M_0$ as the ``bare" mass and write $M(R)$ for 
the ``renormalized" mass, i.e. the resulting mass when $M_0$ is distributed in a spherical shell of 
radius $R$. $M(R)$ takes into account the gravitational self-energy, while $M_0$ corresponds to the sum 
of  all the masses that one would obtain tearing the sphere in many small pieces and moving them  away 
apart. Accordingly, the gravitational potential (\ref{1}) at the surface ($r=R$) should be written as 
\footnote{The equality of inertial and gravitational masses has been tested experimentally even in 
presence of mass defects due to large  binding energies \cite{Sau,Zee}}
\be \Phi(R)=-G\frac{M(R)}{R}
\label{1r}\ee

 How to calculate the renormalized mass $M(R)$ from a given bare mass $M_0$ will be the main point to be 
discussed in the following.

The necessity of taking into account the self-energy, when treating the problem of a black hole, has 
been pointed out recently by Christillin \cite{C}. However his correction is valid only at the first 
order in $c^{-2}$, or, more precisely, at the first order in $R_0/R$, and cannot be used when $U(R)/c^2$ 
is comparable to the bare mass $M_0$. Here we prove that, taking consistently into account the 
implications of the mass-energy equivalence and rewriting (\ref{2}) in terms of the renormalized masses, 
it is impossible to verify (\ref{4}) for any finite $R\neq 0$.  We could say that, while  the 
implementation of special relativity into Newtonian gravitation allows for a ``modern" definition of a 
NBH, on the other side it denies the possibility of its existence.


\section{The consistent mass of a spherical shell and a puzzle}
Given the expression (\ref{6}) for the gravitational energy of a spherical shell, it seems quite natural 
to write down the following consistent equation for the renormalized mass $M(R)$ 
\be M(R)=M_0-\frac{G}{2}\frac{M(R)^2}{Rc^2}
\label{ADM}\ee
whose (positive) solution is
\be M(R)=M_0(-1+\sqrt{1+2R_0/R})R/R_0
\label{sol1}\ee

This equation has been considered since 1960 \cite{ADM,M} in the framework of the classical theory of 
the electron. In fact, adding to (\ref{ADM}) the contribution to the mass of the electromagnetic energy 
$e^2/2R$ (this time positive), the ensuing solution tends to a finite value  when $R\rightarrow 0$: 
$M(R\rightarrow 0)=|e|/\sqrt{G}$, independent of $M_0$. This elegant result exhibits a nice feature of 
the gravitational self-energy as a regularizing device (unfortunately numerically is too big 
($10^{21}m_e$) compared to the electron mass). Instead our interest here is to consider (\ref{ADM}) in 
connection with NBH. From (\ref{sol1}) one sees that $M(R)$ goes to zero for $R\rightarrow 0$ as 
$R^{1/2}$ and that the gravitational potential on the surface of the shell 
\be \Phi(R)=-G\frac{M(R)}{R}=- G\frac{M_0}{R_0}(-1+\sqrt{1+2R_0/R})=-c^2(-1+\sqrt{1+2R_0/R})
\label{phi}\ee
goes to  $-\infty$ for $R\rightarrow 0$.  When $R=2R_0/3$ one gets
 \be\Phi(R=2R_0/3)=-c^2
 \label{phi2/3}\ee
 
Then it seems that taking into account the mass renormalization of the shell, resulting from its 
self-energy, does not prevent the possibility of existence of a NBH; it will only diminish a bit the 
value of the radius at which (\ref{4}) is verified (from $R_0$ to $2R_0/3$).

However there is a contradiction. Suppose we want to deposit a test particle $\delta m_0$ on the surface 
of $M(R)$ and let us think about this test mass as being uniformly distributed on a thin spherical shell 
of radius $r$ centered on the origin, just as $M(R)$. (Note that, neglecting higher orders in $\delta 
m_0$, we do not worry about self-energy of $\delta m_0$ on its own. In other words: $\delta m(r)\approx 
\delta m_0$.) Now imagine to bring $r$ to $R$ and to stick $\delta m_0$ as a thin film on   $M(R)$. 
According to (\ref{3}), if $R=2R_0/3$ the total mass of the system should not increase (or even diminish 
if $R<2R_0/3$), while according to (\ref{sol1}), viewing the system as a new shell of bare mass 
$M_0+\delta m_0$, one has
\be M_t(R)=M(R)+\frac{\partial M(R)}{\partial M_0}\delta m_0= M(R) +\frac{\delta 
m_0}{\sqrt{1+2R_0/R}}>M(R)
\label{Mt1}\ee
in clear contradiction. So there is a mistake somewhere.

We conclude this section with an aside remark.  Analogous considerations hold for an arbitrary spheric 
symmetrical distribution of matter. For instance, in the case of a sphere with uniform volume density, 
one gets a formally identical solution to (\ref{sol1}) with the replacement $R_0\rightarrow 
R_0'=6R_0/5$. The double-shell model that we are exploiting here is most useful since it allows to deal 
with (radial) pointlike particles.

\section{Three recipes for mass renormalization}
In order to discover the origin of the discrepancy we should turn back our attention on how to take into 
account the mutual gravitational interaction energy $U_{int}$ between two bodies of masses $M_1,M_2$. 
Obviously the total mass is 
\be M_{tot}=M_1+M_2+U_{int}/c^2\ee
but how should we split $U_{int}$ between the two bodies? This point is relevant because $U_{int}$, in 
its turn, has to be {\em consistently} expressed  in terms of the modified (fully renormalized) masses. 
To be specific, let us think of $M_1,M_2$ as two pointlike bodies at distance $r$ apart and suppose  
that a fraction $x$ of $U_{int}/c^2$ be attributed to $M_1$ (hence a fraction $1-x$ to $M_2$), then the 
autoconsistent expression for $U_{int}$ will be:
\be U_{int}=-G\frac{(M_1+xU_{int}/c^2)(M_2+(1-x)U_{int}/c^2)}{r}
\label{tb}\ee
which is, in fact, an equation for $U_{int}$ depending on $x$. In \cite{C} it was suggested to attribute 
the whole interaction energy to the smaller mass. Actually in the model at hand we considered two 
concentric shells, the first one with a big mass $M(R)$ (renormalized on its own), the second with an 
infinitesimal mass  $\delta m_0$ that works as a test particle. We thought to stick $\delta m_0$ on the 
surface of the first one, keeping spherical symmetry. In this situation, three possible schemes of 
renormalization are conceivable. In fact the interaction energy between the two shells $\delta U(R)$ may 
be attributed entirely  to the big mass or to the small one, or rather  be split in two equal parts 
between them. In each of these schemes $\delta U(R)$ will assume a specific expression as follows:
\begin{enumerate} 
\item  Renormalization of the big mass $M(R)$:\\
At $1^o$ order in $\delta m_0$  this further renormalization of $M(R)$ can be neglected in $\delta 
U(R)$. So
\be\delta  U(R)=-G\frac{M(R)\delta m_0}{R}
\label{2'}\ee
\item  Renormalization of the small mass $\delta m_0$:\\
In this case  (2) has to be consistently modified, as specified in (\ref{tb})
\be\delta U(R)=-G\frac{M(R)(\delta m_0+\delta U(R)/c^2)}{R}
\label{edU2}\ee
\item  Renormalization of both masses by the same amount: \\
Again at $1^o$ order in $\delta m_0$
\be \delta U(R)=-G\frac{M(R)(\delta m_0+\delta U(R)/2c^2)}{R}
\label{edU3}\ee
\end{enumerate}
The main point comes along  now observing that in the equation for the total mass of the system  \be 
M_t(R)=M(R)+\delta m_0 + \delta U(R)/c^2
\label{tm}\ee
it is
$$M_t(R)-M(R)\equiv dM(R) \quad ;\quad \delta m_0\equiv dM_0$$ 
so that (\ref{tm}) is in fact the differential equation that yields the mass $M(R)$ of a spherical shell 
of radius $R$ as a function of its bare mass $M_0$. Each scheme of renormalization leads to a different 
equation. In the following we shall display the results for each of them.
\vskip15pt
 1. {\em Renormalization of the big mass $M(R)$}\\
Given (\ref{2'}), from (\ref{tm})
\be M_t(R)=M(R)+\delta m_0(1-G\frac{M(R)}{Rc^2})\ee
we get the differential equation
\be \frac{dM(R)}{dM_0}=1- G\frac{M(R)}{Rc^2}
\label{E1}\ee
whose solution is
\be \ln(1-G\frac{M(R)}{Rc^2})=-G\frac{M_0}{Rc^2}
\label{e1}\ee
\be M(R)=\frac{Rc^2}{G}(1-\exp[-G\frac{M_0}{Rc^2}])\equiv M_0(1-\exp[-\frac{R_0}{R}])\frac{R}{R_0}
\label{sol1'}\ee
 Therefore in this scheme, the mass of a spherical shell of radius $R$ and bare mass $M_0$ is not given 
by (\ref{sol1}) (solution of (\ref{ADM})) but by (\ref{sol1'}). The gravitational potential on the 
surface is
\be \Phi(R)=-G\frac{M(R)}{R}=-c^2(1-\exp[-\frac{R_0}{R}])\ee
which keeps finite values and goes to  $-c^2$ only at the limit $R\rightarrow 0$.
 
 \vskip10pt
 2. {\em Renormalization of the small mass $\delta m_0$}\\
 From (\ref{edU2}) 
\be \delta U(R)=-G\frac{M(R)\delta m_0}{R(1+G\frac{M(R)}{Rc^2})} 
\label{dU2}\ee
i.e. the mass $\delta m_0$, once stuck on $M(R)$, is renormalized as
\be \delta m_0 \rightarrow \delta m =\frac{\delta m_0}{1+G\frac{M(R)}{Rc^2}}
\label{dm}\ee
Given (\ref{dU2}), from (\ref{tm})
\be M_t(R)=M(R)+\frac{\delta m_0}{1+G\frac{M(R)}{Rc^2}}\ee
we get the differential equation
\be \frac{dM(R)}{dM_0}=\frac{1}{1+G\frac{M(R)}{Rc^2}}
\label{E2}\ee
whose solution is 
\be M(R)+G\frac{M(R)^2}{2Rc^2}=M_0
\label{e2}\ee
\be M(R)=M_0(-1+\sqrt{1+2R_0/R})R/R_0
\label{sol2}\ee
Here we recover the (\ref{ADM},\ref{sol1}) of Arnowitt, Deser and Missner \cite{ADM}. Now it is clear 
the reason of the inconsistency found above: Using (\ref{sol1}) one should {\em coherently} use 
(\ref{dU2}), not (\ref{2'}). This last equation, for $R=2R_0/3$, would wrongly lead  to $\delta 
U(R)=-c^2\delta m_0$,  instead, according to (\ref{dU2}), it is $\delta U(R=2R_0/3)=-c^2\delta m_0/2$ 
(in agreement with (\ref{Mt1})).

The renormalization of the test mass $\delta m_0$ may be equivalently described in terms of a suitable 
modification of the gravitational potential (for $r>R$)
\be \Phi(r)\rightarrow\Phi_r(r)=-\frac{G}{(1+G\frac{M(R)}{rc^2})}\frac{M(R)}{r}
\label{rgp}\ee
At $r=R$
\be \Phi_r(R)=-c^2(1-\frac{1}{\sqrt{1+2R_0/R}})\ee
Therefore $\Phi_r(R)>-c^2$ for any $R\neq 0$ and $\Phi_r(R)\rightarrow -c^2$ for $R\rightarrow 0$.
\vskip10pt

 3. {\em Renormalization of both masses}\\
 From (\ref{edU3}) we get
 \be \delta U(R)=-G\frac{M(R)\delta m_0}{R(1+G\frac{M(R)}{2Rc^2})} 
\label{dU3}\ee
that means
\be \delta m_0 \rightarrow \delta m =\frac{\delta m_0}{1+G\frac{M(R)}{2Rc^2}}
\label{dm3}\ee
As in the scheme 2, the renormalization of the particle $\delta m_0$ may be equivalently described in 
terms of a suitable modification of the gravitational potential:
\be \Phi(r)\rightarrow\Phi_r(r)=-\frac{G}{(1+G\frac{M(R)}{2rc^2})}\frac{M(R)}{r}
\label{rgp3}\ee
  Given (\ref{dU3}), from (\ref{tm})
\be M_t(R)=M(R)+\frac{1-G\frac{M(R)}{2Rc^2}}{1+G\frac{M(R)}{2Rc^2}}\delta m_0
\label{tm3}\ee
we get the differential equation
\be \frac{dM(R)}{dM_0}=\frac{1-G\frac{M(R)}{2Rc^2}}{1+G\frac{M(R)}{2Rc^2}}
\label{E3}\ee
whose solution will be given by
\be -M(R)-\frac{4Rc^2}{G}\ln(1-\frac{GM(R)}{2Rc^2})=M_0
\label{e3}\ee
Let $z\equiv \frac{GM(R)}{2Rc^2}$, then (\ref{e3}) may be conveniently put as
\be z=1-\exp[-\frac{z}{2}]\exp[-R_0/4R]\approx 
1-\Big(1-\frac{z}{2}+\frac{z^2}{4\cdot 2!}-\cdots \Big)\exp[-R_0/4R] 
\label{e3'}\ee
It is clear from (\ref{e3},\ref{e3'}) that $0\leq z<1$ (for $R\neq 0$), hence (see (\ref{tm3})) 
$M_t(R)>M(R)$. The modified gravitational potential (\ref{rgp3}) at $r=R$ turns out to be
\be \Phi_r(R)=-\frac{2zc^2}{1+z}
\ee
Once again, since $z\rightarrow 1$ for $R\rightarrow 0$,  $\Phi_r(R)\rightarrow -c^2$ at that limit.

\section{Concluding remarks}
 In this paper it was shown that, in a Newtonian theory of gravitation that incorporates the mass-energy 
equivalence, for the interaction energy $\delta U(R)$ between a massive spherically symmetric shell of 
radius $R$ and a test-particle of mass $\delta m_0$, settled on its surface, it is always (for $R\neq 
0$)
\be \delta U(R)>-c^2\delta m_0 
\label{c1}\ee
Here $c^2\delta m_0$ is to be understood as the full relativistic energy of the particle while the 
subscript on the mass indicates that it is ``bare", i.e.  not yet renormalized by the gravitational 
interaction with the heavy shell. The equation $\delta U(R)\leq -c^2\delta m_0 $ was identified as the 
condition for the existence of a Newtonian black hole. Then (\ref{c1}) states that a black hole cannot 
exist in Newtonian gravity.

Besides having taken into account Einstein's mass-energy equivalence (and the  inertial-gravitational 
mass equality), we wrote
\be \delta U(R)=-G\frac{M(R)\delta m}{R}
\label{2''}\ee
where $M(R)$ is the mass of the shell that takes into account its own self-energy (the further 
renormalization of $M(R)$ due to $\delta U(R)$ can be neglected)  and $\delta m$ is the  mass of the 
test particle eventually renormalized by $\delta U(R)$ (the self-energy of $\delta m_0$ on its own can 
be neglected).
 In fact (\ref{c1}) has been established using three possible recipes for mass renormalization that 
differ according to the fraction of the interaction energy $\delta U(R)$ that intervenes in the 
renormalization of $\delta m_0$.

The expression of the renormalized mass $M(R)$ of the shell in terms of its bare  mass $M_0$ is a main 
achievement of the present paper. Depending upon the scheme of renormalization used, we got three 
different solutions that yield rather different results at $R\approx R_0$. However they display the same 
behaviour of $M(R)$ for $R>>R_0$ (i.e. at the first order in $c^{-2}$):
\be M(R)\approx M_0(1-R_0/2R)\ee
and, most remarkably,  the same  lower bound $-c^2\delta m_0$ as regards $\delta U(R)$ with  $\delta 
U(R)\rightarrow -c^2\delta m_0 $ just at the limit $R\rightarrow 0$. This last result is the main one, 
since it implies that in no way a NBH could exist.

 
\vskip20pt
 \noindent {\bf Aknowledgements}
 \vskip10pt
 \noindent The author wish to thank Prof. P. Christillin for friendly assistance  and Prof. G. Morpurgo 
for helpful discussions and  continuous encouragement.

 
\end{document}